\documentclass[english]{bvm2020}
\usepackage{verbatim}
\usepackage{boldline}
\usepackage{graphicx}
\usepackage{caption}
\usepackage{verbatim}
\usepackage{multirow}
\usepackage{xcolor}
\usepackage{boldline}
\usepackage{float}
\usepackage{ulem}
\usepackage{xcolor}

\newcolumntype{L}[1]{>{\raggedright\let\newline\\\arraybackslash\hspace{2pt}}m{#1}}
\newcolumntype{C}[1]{>{\centering\let\newline\\\arraybackslash\vspace{2pt}}m{#1}}
\newcolumntype{R}[1]{>{\raggedleft\let\newline\\\arraybackslash\hspace{2pt}}m{#1}}

\def\articlenumber{0000}

\date{}
\title{Lesson Learnt: Modularization of Deep Networks Allow Cross-Modality Reuse}

\author{Weilin Fu\inst{1,2} \and Lennart Husvogt\inst{1,4} \and Stefan Ploner\inst{1} \and James G. Fujimoto\inst{4} \and Andreas Maier\inst{1,3}}
\authorrunning{F. Author et al.}
\institute{Pattern Recognition Lab, Friedrich-Alexander University\and
International Max Planck Research School Physics of Light (IMPRS-PL)\and
Erlangen Graduate School in Advanced Optical Technologies(SAOT)\and
Biomedical Optical Imaging and Biophotonics Group, MIT, Cambridge, USA
\email{weilin.fu@fau.de}}

\begin{document}

%
\selectlanguage{english}

\maketitle

\begin{abstract}
Fundus photography and Optical Coherence Tomography Angiography (OCT-A) are two commonly used modalities in ophthalmic imaging. With the development of deep learning algorithms, fundus image processing, especially retinal vessel segmentation, has been extensively studied. Built upon the known operator theory, interpretable deep network pipelines with well-defined modules have been constructed on fundus images. In this work, we firstly train a modularized network pipeline for the task of retinal vessel segmentation on the fundus database DRIVE. The pretrained preprocessing module from the pipeline is then directly transferred onto OCT-A data for image quality enhancement without further fine-tuning. Output images show that the preprocessing net can balance the contrast, suppress noise and thereby produce vessel trees with improved connectivity in both image modalities. The visual impression is confirmed by an observer study with five OCT-A experts. Statistics of the grades by the experts indicate that the transferred module improves both the image quality and the diagnostic quality. Our work provides an example that modules within network pipelines that are built upon the known operator theory facilitate cross-modality reuse without additional training or transfer learning.
\end{abstract}

\section{Introduction}
In ophthalmology, fundus photography and optical coherence tomography angiography (OCT-A) are two widely used non-invasive imaging modalities. Fundus photography utilizes fundus cameras to provide 2D RGB images of the retinal surface of the eye, with $30^{\circ}$ to $50^{\circ}$ views of the retinal area at a magnification of $\times2.5$ to $\times5$ times~\cite{srinidhi2017recent}. OCT-A is a 3D imaging technique based on low coherence interferometry, and uses motion contrast to detect blood flow in the retina with micron-scale resolution~\cite{Husvogt2018c,Choi2015}. OCT-A data is often viewed as en face projections which present the 2D view of the retinal vasculature. In both imaging modalities, characterization of the vasculature can strongly support diagnostical procedures. Processing and segmentation of retinal vessels from fundus images is a well-studied field, and several databases with manually labeled pixel-wise vessel annotations have been established~\cite{srinidhi2017recent}. With the recent advances in Deep Learning (DL) technologies, Convolutional Neural Networks (CNNs) are applied on the task and have achieved great success.
However, OCT-A is a modality that has been developed fairly recently, and to the best of our knowledge, there is no vessel segmentation database with manual labels publicly available at the time of writing. This poses difficulties in DL-based algorithms for processing and segmentation of OCT-A data.

Despite that, the resolution of the images and the data distribution of the image intensities are different for the two imaging modalities, there exist structural similarities as presented in Fig.~\ref{fig:fundus} (b) and Fig.~\ref{fig:user_study} (a). Hence it is an instinctive idea to transfer DL-based algorithms which are designed and trained on fundus images to OCT-A data. However, deep networks are in general sensitive to the distribution of the input data, and even intra-modality transfer learning to another database normally requires fine-tuning. In the research direction of Precision Learning~\cite{maier2019learning}, prior knowledge of known operators is incorporated into the CNN architectures to improve the interpretability of the networks. On this basis, a network pipeline composed of two well-defined modules: a preprocessing net and a segmentation net, is constructed for the task of retinal vessel segmentation from fundus images~\cite{fu2019divide}. A small U-Net is employed as the preprocessing net, and Frangi-Net is used to segment the vessels from the processed images. Modularization of the pipeline not only defines specific functions of network blocks, but also allows for flexible reuse of these modules across various tasks as we will show in the following.

In this work, we firstly train the pipeline in~\cite{fu2019divide} for retinal vessel segmentation on the fundus image database DRIVE. Then we use the pretrained preprocessing module directly onto an OCT-A database composed of 20 \mbox{2-D} en face projection images. Due to the absence of ground truth data with clear vessels and clean background, an observer study based on five datasets involving five OCT-A experts is conducted. Feedback from the experts suggests that the images prepared with the pretrained preprocessing module have less noise and improved vessel network connectivity, and can thus potentially better assist the diagnosis procedure. This result indicates that the preprocessing module retains its edge-preserving denoising ability and is reusable across different imaging modalities.

\section{Materials and Methods}
\subsection{Preprocessing Network}
\begin{figure}[t]
    \centering
    \begin{minipage}{0.95\textwidth}
    \subfigure
    {\includegraphics[width=\textwidth]{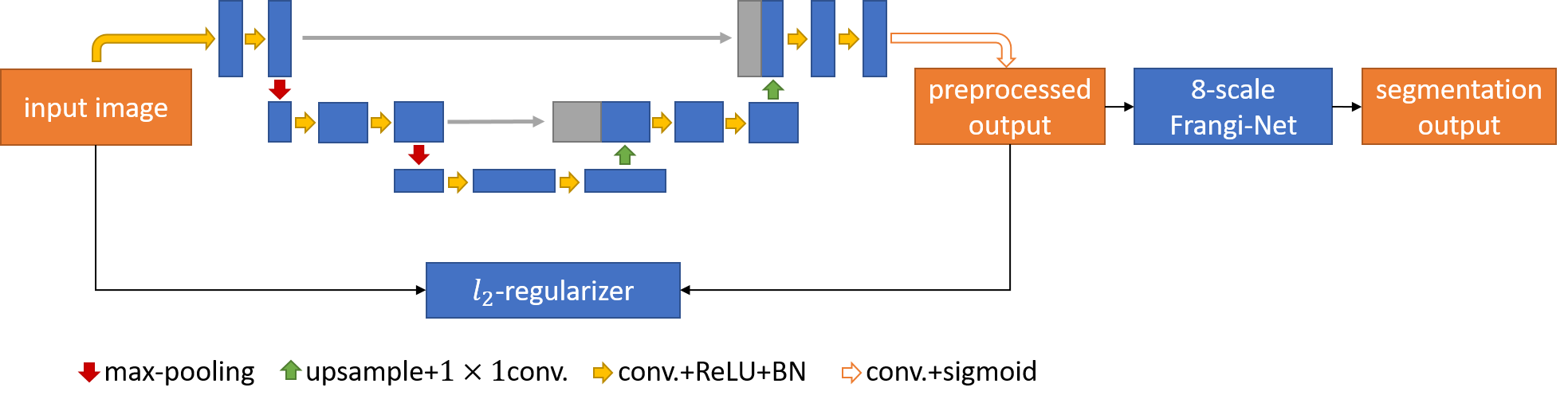}}
    \end{minipage}
    \caption{The architecture of the retinal vessel segmentation network on fundus images. The preprocessing module is the U-Net on the left.}\label{fig:pretrain}
\end{figure}

\textcolor{black}{The preprocessing module is adopted from the network pipeline in~\cite{fu2019divide}, as shown in Fig.~\ref{fig:pretrain}. In this workflow, a three-level U-Net with 16 filters in the input convolutional layer is employed as the preprocessing net, and an eight-scale Frangi-Net is used for segmentation. In the preprocessing part, a Mean Square Error (MSE) regularizer is utilized to constrain the similarity between the input image and the preprocessed output. The overall pipeline is trained end-to-end on the fundus image database DRIVE as a retinal vessel segmentation task.}

\textcolor{black}{Each training batch contains 50 patches of size $168\times168$ pixels. Data augmentation techniques such as additive noise, rotation and scaling are employed for better generalization. The objective function consists of three main parts: weighted focal loss~\cite{lin2017focal}, $\ell_2$-norm to confine the weights in U-Net, and the MSE similarity regularizer. Optimization is performed with Adam optimizer~\cite{kingma2014adam}. The learning rate is initialized to $5\times10^{-5}$ and decays after each 10k steps. Early stopping is applied according to the validation loss curve.}

\subsection{Reader Study}
\textcolor{black}{Retrospective data assessment by five experts is used for this study. In each experiment, three images are presented in random order, namely the raw OCT-A en face projection image, the output from the preprocessing net, and a blend of these two (50\,\% each). The experts are requested to grade the images from 1 (very good) to 5 (very bad) with respect to three aspects: image quality regarding to the noise level, vessel connectivity and the diagnosis quality. The observers are allowed to adjust the brightness and contrast of the given images. The mean score of each image type on each quality aspect over all experiments, as well as the corresponding inter-expert standard deviation are reported.}

\subsection{Database Description}
\subsubsection{Fundus Training Database:}
\begin{figure}[t]
    \centering
    \begin{minipage}{0.3\textwidth}
    \subfigure[]
    {\includegraphics[width=\textwidth]{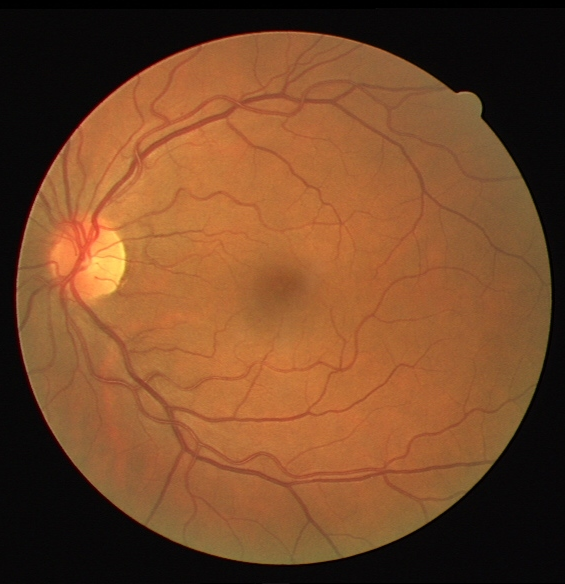}}
    \end{minipage}
    \hfill
    \begin{minipage}{0.3\textwidth}
    \subfigure[]
    {\includegraphics[width=\textwidth]{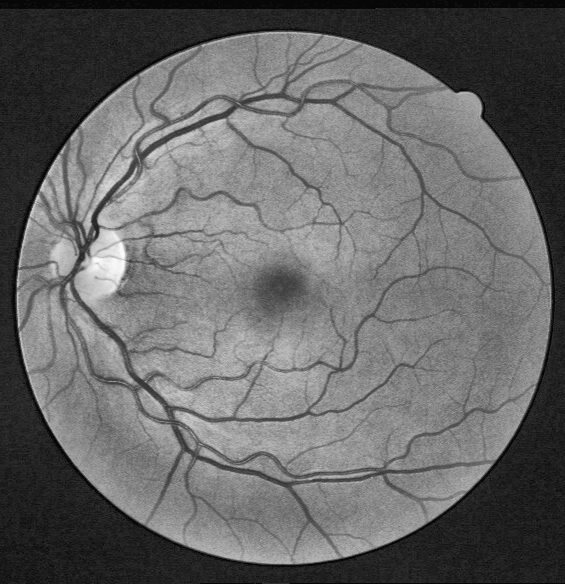}}
    \end{minipage}
    \hfill
    \begin{minipage}{0.3\textwidth}
    \subfigure[]
    {\includegraphics[width=\textwidth]{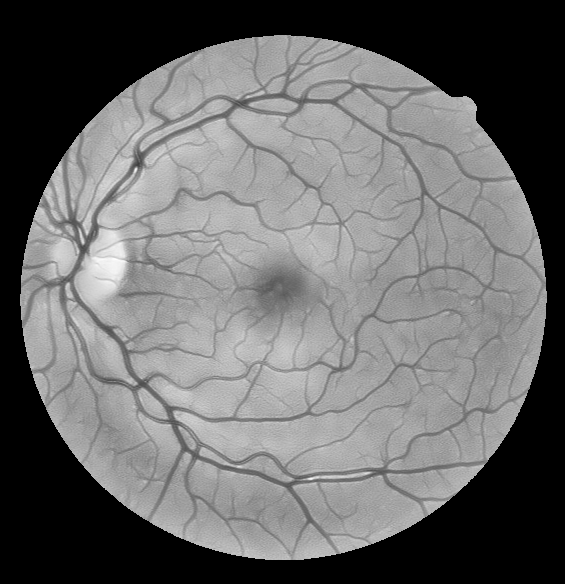}}
    \end{minipage}
    \caption{The raw fundus image in (a). Input and output of the preprocessing network in (b) and (c), respectively. Example image is test01.TIFF from the DRIVE database.}\label{fig:fundus}
\end{figure}

\textcolor{black}{The Digital Retinal Images for Vessel Extraction (DRIVE) database which contains 40 RGB fundus images is used for training the network pipeline as a vessel segmentation task. All images in DRIVE are of size $565\times 584$ pixels, and are provided with manual labels and Field of View (FOV) masks. The raw images are prepared with the pipeline of green channel extraction, illumination balance with CLAHE~\cite{zuiderveld1994contrast}, and intensity standardization to (-1, 1). Note that the intensity of regions where vessel diameters are below 8 pixels normally have intensities between (-0.6, 0.6). The database is equally divided into one training and testing set, and a validation set containing four images is separated from the training set. During the training progress, a multiplicative pixel-wise weight map which is inversely proportional to the ground truth vessel diameter is generated for each image to emphasize on thin vessels. One representation from DRIVE is presented in Fig.~\ref{fig:fundus} (a), the corresponding input and output of the preprocessing net are shown in Fig.~\ref{fig:fundus} (b)-(c).}

\subsubsection{OCT-A Testing Database:} 
\textcolor{black}{The testing OCT-A data in this study are acquired from a healthy 28-year-old male volunteer with an ultrahigh speed swept source OCT research prototype developed at the Massachusetts Institute of Technology and used by the New England Eye Center at Tufts Medical Center in Boston~\cite{Choi2015}. The database contains 20 en face OCT-A images of size $500\times500$ pixels, where 10 images have the field size of $3\times3\,\textrm{mm}$ and the other 10 images have the field size of $6\times 6\,\textrm{mm}$. Contrary to those in fundus images, vessels in OCT-A are represented as bright tubular structures in dark background. The pixel intensities of capillary regions range from 0 to around 1.5. To adjust the data range of the testing databases, the following linear intensity transform is applied on the OCT-A database:
Firstly a threshold $4.0$ is set, since the contrast in the big bright vessels are not of interest in this work. The images are inverted by multiplying $-1$ and then added with $0.5$ such that the intensities in small vessel regions roughly match that in fundus images.}

\begin{figure}[t]
    \centering
    \begin{minipage}{0.3\textwidth}
    \subfigure[]
    {\includegraphics[width=\textwidth]{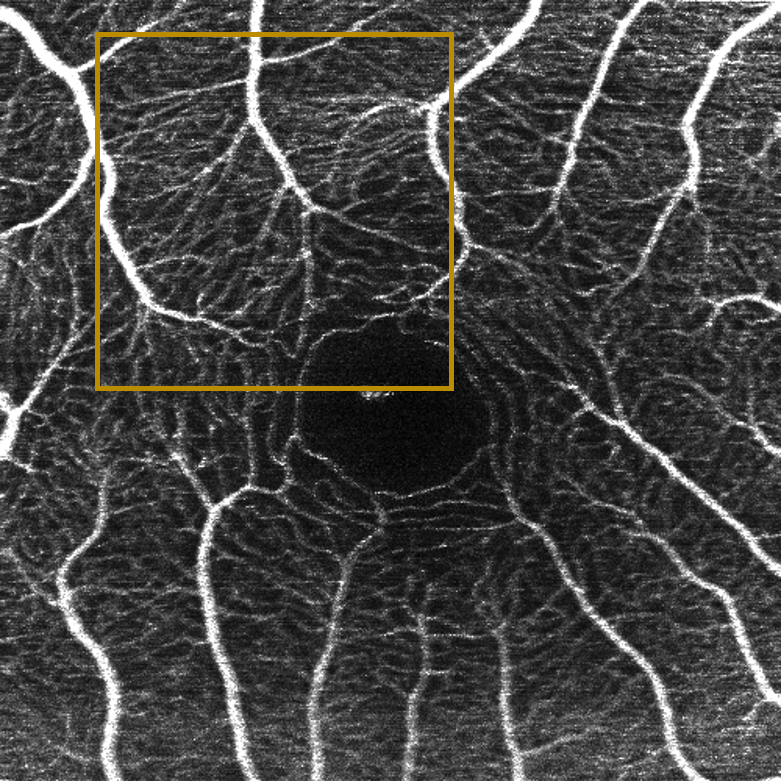}}
    \end{minipage}
    \hfill
    \begin{minipage}{0.3\textwidth}
    \subfigure[]
    {\includegraphics[width=\textwidth]{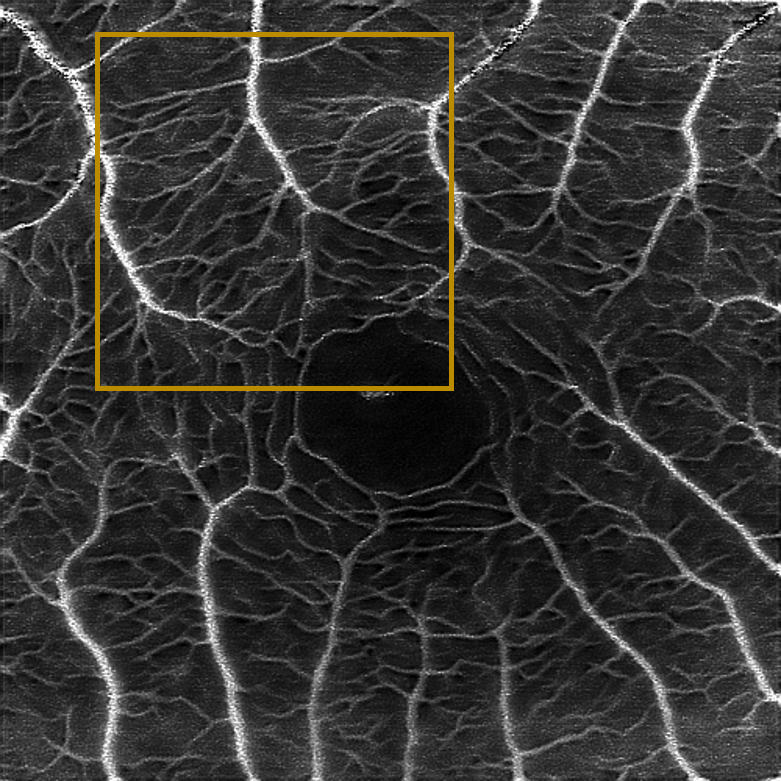}}
    \end{minipage}
    \hfill
    \begin{minipage}{0.3\textwidth}
    \subfigure[]
    {\includegraphics[width=\textwidth]{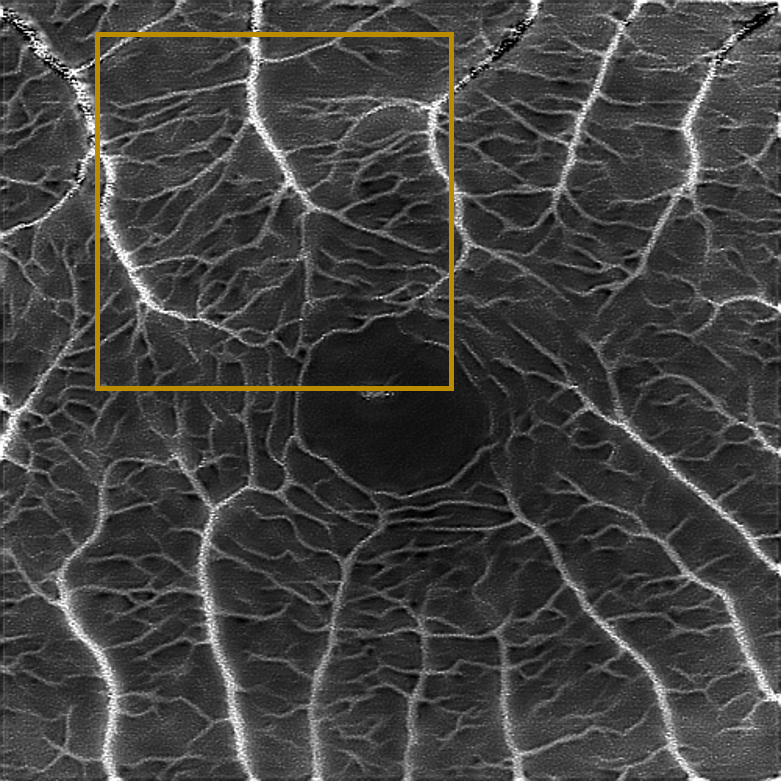}}
    \end{minipage}
    \\
    \begin{minipage}{0.3\textwidth}
    \subfigure[]
    {\includegraphics[width=\textwidth]{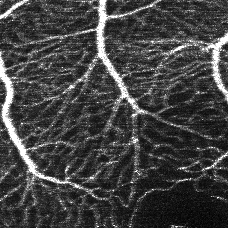}}
    \end{minipage}
    \hfill
    \begin{minipage}{0.3\textwidth}
    \subfigure[]
    {\includegraphics[width=\textwidth]{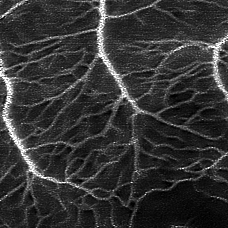}}
    \end{minipage}
    \hfill
    \begin{minipage}{0.3\textwidth}
    \subfigure[]
    {\includegraphics[width=\textwidth]{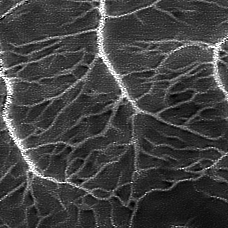}}
    \end{minipage}
    \caption{One example in the OCT-A database. The original image and ROI in the yellow box shown in (a) and (d). The outputs from the preprocessing net in (c) and (f). The blend images in (b) and (c). Field size is $3\times 3\,\textrm{mm}$ in the original image.}\label{fig:user_study}
\end{figure}

\section{Results}

\textcolor{black}{Images in one representative experiment with enlarged Regions Of Interest (ROIs) are presented in Fig.~\ref{fig:user_study}. Direct visual impact indicates that in the preprocessed and the blend images, the noise level is reduced and the vascular structures are enhanced. These changes introduce cleaner boundaries and better-connected vessels. However, not all emerged vessels can be visually validated according to the given raw OCT-A image, i.e. some could be hallucinated by the preprocessing net. In addition, the high intensities within the thick vessels can be out of data range for the network and thus cause black responses in the output. Blending of the raw image and the output of the preprocessing net could mitigate these issues.}
\textcolor{black}{The visual impression of the three image types is confirmed with the statistical results of the observer study. Despite of the subjective influence which can be reflected by the inter-expert standard deviation, the output and the blend images achieve better scores than the raw input with respect to image quality, vessel connectivity as well as potential diagnosis quality, as shown in Table~\ref{tab:user}.}

\begin{table}[]
\centering
\caption{The mean and standard deviation of the observer study. IQ, VC, DQ refer to Image Quality regarding to noise level, Vessel Connectivity, and Diagnoistic Quality, respectively. The grades range from 1 (very good) to 5 (very bad).}
\label{tab:user}
\begin{tabular}{C{1.5cm}C{2.5cm}C{2.5cm}C{2.5cm}} \clineB{1-4}{2.}
\textbf{} & \multicolumn{1}{c}{\textbf{raw input}} & \textbf{blend}   & \textbf{output}    \\ \hline
IQ                & $3.0\pm 0.8$                               & $2.2\pm0.6$   & $2.2\pm0.3$  \\
VC                & $3.1\pm 0.7$                               & $2.1\pm0.5$   & $2.2\pm0.7$  \\
DQ                & $3.0\pm 0.8$                               & $2.0\pm0.6$   & $2.2\pm0.5$  \\
\clineB{1-4}{2.}
\end{tabular}
\end{table}

\section{Discussion}
\textcolor{black}{In this work, we transfer a preprocessing net which is pretrained on the fundus database DRIVE directly onto OCT-A en face projection images without further training. Direct visual inspection and an observer study indicate that the preprocessing network notably enhances the OCT-A images regarding to image quality, vessel connectivity and potential diagnosis quality.} \textcolor{black}{To the best of our knowledge, this is the first work of cross-modality CNN module transfer without further network fine-tuning or transfer learning.} \textcolor{black}{Despite the difference in input data distribution, the network performs a similar function on both modalities: balancing the contrast, reducing the noise level, and improving the vessel connectivity. This work provides one example of the successful reuse of modules within CNN pipelines which are constructed according to the known operator theory.}

\textcolor{black}{In the future, the image quality enhancement by the transferred net will be quantitatively validated with reconstructed high-resolution image of the OCT-A data.}
\textcolor{black}{The preprocessing module could also be reused to improve the image quality in different data modalities. As an extension, pretrained modules from different network pipelines could be recombined for new tasks.}
\textcolor{black}{ Finally, the network could also be incorporated into an OCT-A reconstruction pipeline based on compressed sensing, where it could serve as a regularizer~\cite{Husvogt2019}.}

\section{Acknowledgements}
The research leading to these results has received funding from the European Research Council (ERC) under the European Union’s Horizon 2020 research and innovation programme (ERC grant no. 810316).

\bibliographystyle{bvm2020}

\bibliography{0000}

\begin{thebibliography}{1}

\bibitem{srinidhi2017recent}
Srinidhi CL, Aparna P, Rajan J.
\newblock {Recent Advancements in Retinal Vessel Segmentation}.
\newblock Journal of medical systems. 2017;41(4):70.

\bibitem{Husvogt2018c}
Husvogt L, Ploner S, Maier A.
\newblock {Optical Coherence Tomography}.
\newblock Springer, Cham; 2018.  p. 251--261.

\bibitem{Choi2015}
Choi W, Moult EM, Waheed NK, et~al.
\newblock {Ultrahigh-Speed, Swept-Source Optical Coherence Tomography
  Angiography in Nonexudative Age-Related Macular Degeneration with Geographic
  Atrophy}.
\newblock Ophthalmology. 2015;.

\bibitem{maier2019learning}
Maier AK, Syben C, Stimpel B, et~al.
\newblock {Learning with Known Operators Reduces Maximum Error Bounds}.
\newblock Nature machine intelligence. 2019;1(8):373--380.

\bibitem{fu2019divide}
Fu W, Breininger K, Schaffert R, et~al.
\newblock {A Divide-And-Conquer Approach Towards Understanding Deep Networks}.
\newblock In: MICCAI. Springer; 2019.  p. 183--191.

\bibitem{lin2017focal}
Lin TY, Goyal P, Girshick R, et~al.
\newblock {Focal Loss for Dense Object Detection}.
\newblock In: Proceedings of the IEEE international conference on computer
  vision; 2017.  p. 2980--2988.

\bibitem{kingma2014adam}
Kingma DP, Ba J.
\newblock {Adam: A Method for Stochastic Optimization}.
\newblock arXiv preprint arXiv:14126980. 2014;.

\bibitem{zuiderveld1994contrast}
Zuiderveld K.
\newblock {Contrast Limited Adaptive Histogram Equalization}.
\newblock In: Graphics gems IV. Academic Press Professional, Inc.; 1994.  p.
  474--485.

\bibitem{Husvogt2019}
Husvogt L, Ploner S, Moult EM, et~al.
\newblock {Using Medical Image Reconstruction Methods for Denoising of OCTA
  Data}.
\newblock In: Investigative Ophthalmology {\&} Visual Science. vol.~60.
  {Association for Research in Vision and Ophthalmology}; 2019.  p. 3096.

\end{thebibliography}
\marginpar{\color{white}E\articlenumber} 
\end{document}